\documentclass[twocolumn]{aastex631}

\usepackage{amsmath}
\usepackage{appendix}
\definecolor{blue-violet}{rgb}{0.30, 0.1, 0.89}

\begin{document}

\title{Relativistic Jets and Winds in Radio-Identified Supermassive Black Hole Binary Candidates}

\author[0000-0002-9545-7286]{Andrew G. Sullivan}
\affiliation{Kavli Institute for Particle Astrophysics and Cosmology, Department of Physics, Stanford University, Stanford, CA 94305, USA}
\author[0000-0002-1854-5506]{Roger D. Blandford}
\affiliation{Kavli Institute for Particle Astrophysics and Cosmology, Department of Physics,
Stanford University, Stanford, CA 94305, USA}

\author[0009-0004-2614-830X]{Anna Synani} 
\affiliation{Department of Physics and Institute of Theoretical and Computational Physics, University of Crete, 71003 Heraklion, Greece}
\affiliation{Institute of Astrophysics, Foundation for Research and Technology-Hellas, GR-71110 Heraklion, Greece}

\author[0000-0001-5957-1412]{Philipe V.~de la Parra}
\affiliation{\hbox{CePIA}, Astronomy Department, Universidad de Concepci\'on,  Casilla~\hbox{160-C}, Concepci\'on, Chile}

\author[0000-0001-9011-0737]{No\'emie Globus}
\affiliation{Instituto de Astronom{\'\i}a, Universidad Nacional Aut\'onoma de M\'exico, km 107 Carretera Tijuana-Ensenada, 22860 Ensenada, Baja California, México}

\author[0000-0003-0936-8488]{Mitchell C. Begelman}
\affiliation{JILA, University of Colorado and National Institute of Standards and Technology, 440 UCB, Boulder, CO 80309-0440, USA}
\affiliation{Department of Astrophysical and Planetary Sciences, University of Colorado, 391 UCB, Boulder, CO 80309-0391, USA}


\author[0000-0001-9152-961X]{Anthony C.S. Readhead}
\affiliation{Owens Valley Radio Observatory, California Institute of Technology,  Pasadena, CA 91125, USA}



\begin{abstract}
Supermassive black hole binary systems (SMBHBs) are thought to emit the recently discovered nHz gravitational wave background; however, not a single individual nHz source has been confirmed to date. Long-term radio-monitoring at the Owens Valley Radio Observatory has revealed two potential SMBHB candidates: blazars PKS 2131-021 and PKS J0805-0111. These sources show periodic flux density variations across the electromagnetic spectrum, signaling the presence of a good clock.   
To explain the emission, we propose a generalizable jet model, where a mildly relativistic wind creates an outward-moving helical channel, along which the ultra-relativistic jet propagates. The observed flux variation from the jet is mostly due to aberration. The emission at lower frequency arises at larger radius and its variation is consequently delayed, as observed.
Our model reproduces the main observable features of both sources and can be applied to other sources as they are discovered. We make predictions for radio polarization, direct imaging, and emission line variation, which can be tested with forthcoming observations. Our results motivate future numerical simulations of jetted SMBHB systems and have implications for the fueling, structure, and evolution of blazar jets. 

\end{abstract}


\section{Introduction} \label{sec:intro}
The discovery of a nHz gravitational wave (GW) background by the global network of pulsar timing arrays (PTAs) provides strong evidence for the ubiquity of supermassive black hole binaries (SMBHBs) throughout the universe \citep{2019A&ARv..27....5B, 2023ApJ...951L...8A, 2023ApJ...952L..37A, 2023A&A...678A..50E, 2023ApJ...951L...6R, 2023RAA....23g5024X}. Direct evidence for such binaries includes 4C+37.11 \citep{2004ApJ...602..123M,2014ApJ...780..149R, 2017ApJ...843...14B, 2024ApJ...960..110S}, whose black holes have projected separation $7.3$ pc, as well as the quasi-periodic blazar OJ 287 \citep{1988ApJ...325..628S,2022MNRAS.513.3165K}.  
However, despite the promise of searching for periodicity in active galactic nuclei (AGN) light curves as a discovery method, statistically strong SMBHB candidates have proven elusive \citep{2016MNRAS.461.3145V, 2019ApJ...884...36L}.

The situation has recently changed, with the discovery of two SMBHB candidate blazars PKS 2131-021 (hereafter PKS 2131) and PKS J0805-0111 (hereafter PKS J0805). Both show strong statistical evidence for periodic variability. For PKS 2131, periodic variation re-appeared after a 19-yr hiatus with the same period and phase (see below). With periods of only a few years, these candidates may contribute to the nHz GW background. 

PKS 2131, a BL Lac object at $z_S=1.283$, has shown coherent $P=1739.8\pm17.4$ day periodicity over 47.9 years in radio \citep{2021MNRAS.506.3791R,2022ApJ...926L..35O, 2024Paper2}. PKS J0805, a flat-spectrum radio quasar (FSRQ) at redshift $z_S=1.388$ \citep{2007ApJS..171...61H, 2008ApJS..175...97H}, has shown a $P=1240.7\pm4.6$ day periodicity over 14 years of radio observation \citep{2021RAA....21...75R,2025ApJ...987..191D}. Recent observations have revealed the same periodicities across the spectrum, from cm to optical wavelengths, and possibly even gamma-rays \citep{2024Paper2, 2025arXiv250404278H}.

Motivated by the observations of PKS 2131 and PKS J0805, we present a model for jetted SMBHB systems. We invoke a sub-relativistic wind to collimate the jet launched from one of the two black holes in the binary. The wind, dragged around by the orbital motion, creates a conical-helical channel through which the relativistic flow travels, producing sinusoidal aberration.\footnote{Note that, at radii much greater than the orbital radius, it is the orbital velocity that determines the wind direction. The jet direction consequently varies in quadrature with the displacement and so the jet kinematics is similar to that induced by jet precession.} This work enhances the simple model presented in \citet{2022ApJ...926L..35O} and \citet{2024Paper2},  in which the relativistic jet propagates ballistically, requiring coherent variation over $\sim100$ pc scales. We detail the model in Sec.~\ref{sec:helicaljet}.  We interpret the observations of PKS 2131 and PKS J0805 and apply our model in Sec.~\ref{sec:SMBHbinary}. In Sec.~\ref{sec:futureobs}, we consider the prospects of future emission line and GW observations, before concluding in Sec.~\ref{sec:conclusion}.

\section{Flux variation from Disk-wind Collimated Jets}
\label{sec:helicaljet}

\begin{figure}
    \centering
    \includegraphics[width=\linewidth]{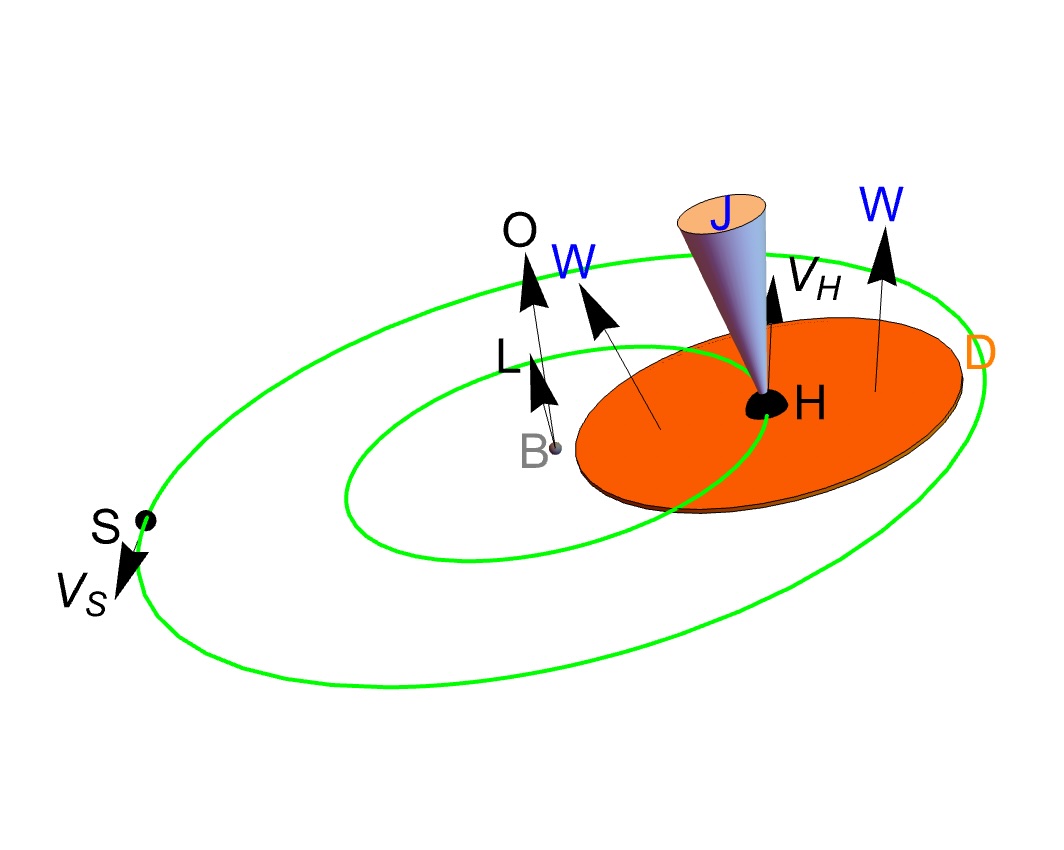}
    \includegraphics[width=\linewidth]{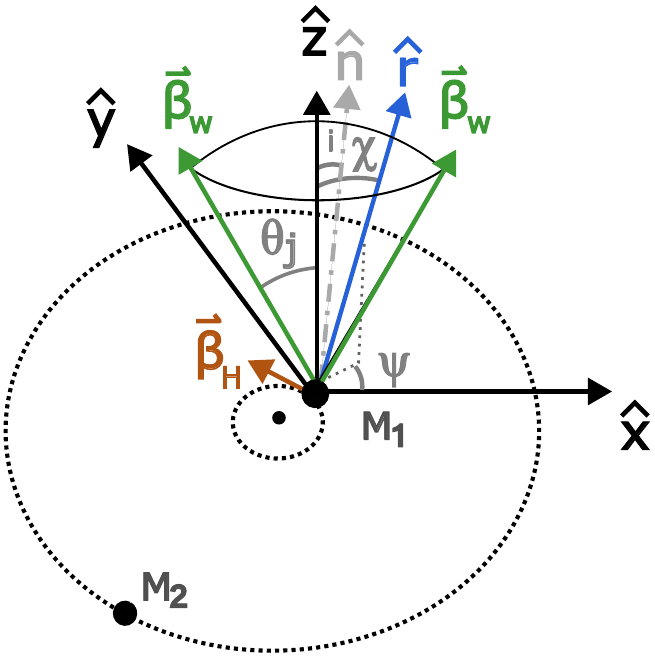}
    \includegraphics[width=1\linewidth]{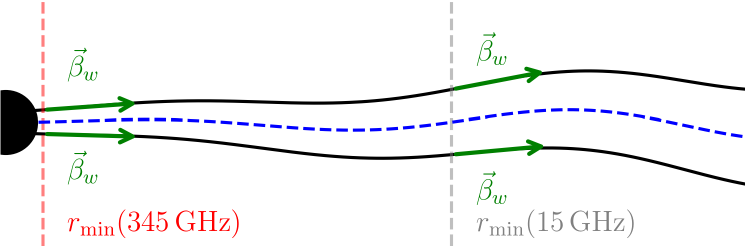}

    \caption{A visualization of our model. (Top) An SMBHB composed of the primary black hole H and a secondary black hole S. H orbits with speed $v_H$, carries around a disk (D) and launches a jet (J) along the orbital angular momentum direction (L). The disk produces a wind (W) with speed $v_w<c$, that collimates the jet. (Middle) A geometric illustration of the conical jet along the $\hat{z}$ axis with opening angle $\theta_j$ confined by the wind. The wind-jet interface has velocity $\vec{\beta}_w=\vec{v}_w/c$. We show a unit vector $\hat{r}$ associated with angular position ($\chi$, $\psi$) as well as the observer viewing direction $\hat{n}$. The orbital motion with velocity $\vec{\beta}_H=\vec{v}_H/c$ will distort the jet into a conical helix. (Bottom) A 2D slice through the wind-collimated helical jet. The dashed line in the center denotes the sinuous path of relativistic flow moving through the center of the channel. The minimum radii $r_{\min}$ which contribute at two frequencies are labeled. }
    \label{fig:modeldrawing}
\end{figure}
In our model, the SMBHB has period $P$ and is composed of primary mass $M_1$ and secondary mass $M_2$. We envision one black hole (either the primary or secondary) launching a jet. If the other black hole also launches a jet, we can add a second jet whose orbital motion is half a period out of phase. The main observational feature will be additional higher harmonics. We imagine a sub-relativistic collimating wind, which confines the synchrotron emitting relativistic jet with velocity $\beta_j$ and associated bulk Lorentz factor $\Gamma_j$. We suppose the interface between the jet and wind has effective speed $\beta_w \lesssim0.9$ (i.e. lying between the sub-relativistic wind and highly relativistic jet speeds). In this model, the wind plays a different role from the sheath in \cite{2017MNRAS.465..161S} because it does not contribute to the radiation. The wind serves primarily to confine the jet, while the radiation comes from inside the jet. 
 
The sub-relativistic wind may be a hydromagnetic flow produced from an extended disk \citep[e.g.][]{1982MNRAS.199..883B,1992ApJ...385..460E, 2016MNRAS.461.2605G} or a jet sheath \citep[e.g.][]{1989A&A...224...24P, 1989MNRAS.237..411S, 2017MNRAS.465..161S}. 
We assume the inertia of the wind-jet complex is predominantly carried by the wind.
Consequently, we imagine that the relativistic jet direction is completely determined by the wind-jet interface as in the middle panel of Fig.~\ref{fig:modeldrawing}. The orbital motion will add to the effective wind velocity, distorting the direction sinusoidally, forming a conical helix as in the bottom panel of Fig.~\ref{fig:modeldrawing}. Functionally, this model is similar to those presented in \citet{2017MNRAS.465..161S} and \citet{2022ApJ...926L..35O}, except that the helix shape and propagation speed are set by the wind-jet interface rather than a ballistically propagating jet.

In the binary center-of-mass frame, the primary moves with velocity $\vec{\beta}_H=\vec{v}_H/{c}$ (See the top panel of Fig. \ref{fig:modeldrawing}). $\beta_H$ is related to the GW radiation lifetime $\tau$
\begin{equation}
    \beta_H=0.31\left(\frac{\tau}{P}\right)^{-\frac{1}{5}}\frac{q^{\frac{4}{5}}}{\left(1+q\right)^{3/5}},
\end{equation}
where $q=M_2/M_1$ is the mass ratio. A reasonable binary lifetime $\tau\gtrsim10^4\,P$
would suggest $\beta_H\lesssim0.02<\beta_w$ (with $q\approx0.5$). Since the orbital motion modulates the direction of the collimating wind and thus the confined relativistic flow velocity, the jet Doppler factor $D$ changes relative to the observer. 

The inclusion of the wind in our model adds a lever on the emission region locations at different frequencies. If the jet were to propagate ballistically with $\Gamma_j$, phase offsets in the observed sinusoidal light curves at different observing frequencies seen in PKS 2131 and PKS J0805 would suggest emission regions separated by $\sim100$ pc \citep{2024Paper2}. These large separations would require the helical structure to remain coherent over $\gtrsim100$ helical wavelengths $\lambda_h=\beta_wcP$ (where $\beta_w=\beta_j$ if the relativistic flow is ballistic). With the jet channel propagating at a non-ultra-relativistic $\beta_w$, the emission regions can be separated by $\lesssim10\lambda_h$.

\subsection{Basic Jet Model}
We suppose that the jet is confined within a cone of constant opening angle $\theta_j$. The observer direction makes an angle $i$ with the cone axis. The jet velocity is radial and constant throughout the jet, so that $\vec \beta_j=\beta_j \hat{r}$, where $\hat{r}$ is a radial unit vector associated with a particular angular position within the open jet (see middle panel of Fig.~\ref{fig:modeldrawing}).  
We assume that the isotropic relativistic electron distribution function in the comoving jet (primed) frame is $n^\prime(\gamma)=K^\prime\gamma^{-p}$ for $\gamma_{e,\rm min}<\gamma<\gamma_{e,\rm max}$. The associated electron pressure is $P^\prime_e=\frac{1}{3}K^\prime m_e c^2 \int_{\gamma_{e,\mathrm min}}^{\gamma_{e,\mathrm max}}\gamma^{1-p}d\gamma$, where $m_e$ is electron mass. We normalize the pressure using the power in relativistic electrons $L_e=4\pi P^\prime_e(\Gamma_j\theta_jr)^2c$, where $r$ is radius from the black hole. Likewise, the comoving magnetic field strength $B^\prime$, assumed to be mostly transverse and disordered, is normalized by the electromagnetic power $L_m=(B^\prime\Gamma_j\theta_jr)^2c/4$.

The synchrotron emissivity in the lab frame at {lab frame} frequency $\nu$ and {angular position $\Omega$} will be
\begin{equation}
\begin{split}
\label{eq:emissivity}
j_{\nu,\Omega}  =&D^{\frac{3+p}{2}}j^\prime_{\nu,\Omega}\\=&D^{\frac{3+p}{2}}\frac{\sqrt{3\pi}}{16\pi^2 m_e c^2}q_e^3 \frac{K^\prime}{p+1}\left(\frac{2\pi m_ec}{3q_e}\right)^{-\frac{p-1}{2}}B^{\prime \frac{p+1}{2}}\\&\times {\mathrm {\Gamma}}\left(\frac{p}{4}+\frac{19}{12}\right){ {\Gamma}}\left(\frac{p}{4}-\frac{1}{12}\right) \\&\times { {\Gamma}}\left(\frac{p}{4}+\frac{5}{4}\right){\Gamma}\left(\frac{p}{4}+\frac{7}{4}\right)^{-1}\nu^{-\alpha},
\end{split}
\end{equation}
where $j^\prime_{\nu,\Omega}$ is the {pitch angle-averaged} emissivity in the comoving jet frame, $\Gamma$ denotes the Gamma function (not to be confused with jet bulk Lorentz factor $\Gamma_j$), $q_e$ here is electron charge, and $\alpha=(p-1)/2$ is the spectral index \citep{1979rpa..book.....R}. 
The absorption coefficient in the observer frame is
\begin{equation}
\label{eq:absorption}
\begin{split}
\mu_\nu=&D^{\frac{p+2}{2}} \mu^\prime_\nu \\=&D^{\frac{p+2}{2}} \frac{\sqrt{3\pi}}{32\pi^2 m_e^2 c^2} \left(\frac{3q_e}{2\pi m_e c}\right)^{\frac{p}{2}} K^\prime  B^{\prime \frac{p+2}{2}}\\&\times\Gamma\left(\frac{p}{4}+\frac{22}{12}\right)\Gamma\left(\frac{p}{4}+\frac{2}{12}\right)\\&\times \Gamma\left(\frac{p}{4}+\frac{6}{4}\right)\Gamma\left(\frac{p}{4}+\frac{8}{4}\right)^{-1}\nu^{-\frac{p+4}{2}},
\end{split}
\end{equation}
where $\mu_\nu^\prime$ denotes the absorption coefficient in the flow frame.
For radio emitting particles in $B\sim10$ mG fields, the synchrotron cooling times are much longer than the orbital period, so adiabatic losses dominate in the cm emission region. This may not be the case for mm emission regions where the fields may reach up to 1 G, but we leave the inclusion of cooling to future work. {For this work, we assume one sustained particle spectrum across the radio emission regions without cooling, which corresponds to assuming a volume-filling acceleration zone with very fast acceleration relative to cooling.} We return to acceleration mechanisms in Sec. \ref{sec:PKS2131_model}.

Orbital motion causes major variation in $D$ and consequently flux density variation due to coherent aberration across the jet. To allow coherent helical motion, hydromagnetic waves must be able to cross the jet on a timescale comparable to the {flow time through the jet helix} (i.e. $\Gamma_j\theta_j\lesssim 1/\beta_w$). In practice, the jet kinematics is likely significantly more complicated and variable. We discuss some of these complications in Sec. \ref{sec:conclusion}. The underlying ``clock'' should not exhibit frequency noise. 

\subsection{Kinematics}
Due to the orbital motion and collimation by the wind, a radial unit vector $\hat{r}$ in a particular direction within the open jet will be redirected 
\begin{equation}
\label{eq:direction}
    \hat{r} =\frac{1}{\sqrt{1+2\hat{r}_0\cdot\frac{\vec{\beta}_H}{\beta_w}+\frac{\beta_H^2}{\beta_w^2}}}\left(\hat{r}_0+ \frac{\vec{\beta}_H}{\beta_w}\right),
\end{equation}
where $\hat{r}_0$ is the unperturbed radial unit vector, which corresponds to a fixed direction in the black hole frame. Note that $\vec{\beta}_H$ will introduce a sinusoidal perturbation to $\hat{r}$. At a particular radius $r$ from the black hole, $\vec{\beta}_H$ in eq. \ref{eq:direction} is delayed in phase from the instantaneous orbital velocity by $r/\lambda_h$.

The local relativistic flow velocity in the observer frame is 
\begin{equation}
\label{eq:betaj}
    \vec{\beta}_j = \beta_j \hat{r} \approx \beta_j\left(\hat{r}_0\left(1- \hat{r}_0\cdot \frac{\vec{\beta}{_H}}{\beta_w}\right)+\frac{\vec{\beta}_H}{\beta_w} \right).
\end{equation}
The jet Doppler factor is 
\begin{equation}
\label{eq:Doppler}
    D=\frac{1}{\Gamma_j\left(1-\beta_j\cos\theta\right)},
\end{equation}
where $\cos\theta=\hat{r}\cdot\hat{n}$ and $\hat{n}$ is the fixed observer direction. We expand the Doppler factor
\begin{equation}
    D=D(\cos\theta_0)+\frac{\partial D}{\partial \cos\theta}(\cos\theta_0) \delta\cos\theta,
\end{equation}
where 
\begin{equation}
    \frac{\partial D}{\partial\cos\theta}=\frac{\beta_j}{\Gamma_j\left(1-\beta_j\cos\theta\right)^2},
\end{equation}
and
\begin{equation}
    \delta\cos\theta=\left(\hat{n}-\left(\hat{r}_0\cdot\hat{n}\right)\hat{r}_0\right)\cdot\frac{\vec{\beta}_H}{\beta_w}.
\end{equation}
Note that $\cos \theta_0=\hat{r}_0\cdot\hat{n}$. Therefore, the variation in the Doppler factor is 
\begin{equation}
    \delta D = \frac{\beta_j\left(\hat{n}-\left(\hat{r}_0\cdot\hat{n}\right)\hat{r}_0\right)}{\Gamma_j\left(1-\beta_j(\hat{r}_0\cdot\hat{n})\right)^2}\cdot\frac{\vec{\beta}_H}{\beta_w}.
\end{equation}

\subsubsection{Amplitude of Flux Density Variations}
 \label{sec:OrbitalKinematicModel}
We now define an angular position within the jet ($\chi$, $\psi$) where $\chi$ is the polar angle off the jet axis (assumed to be along $\hat{z}$), and $\psi$ is the azimuthal angle. In terms of $\chi$ and $\psi$, $\hat{r}_0$ is
\begin{equation}
  \hat{r}_0=\cos\psi \sin\chi \hat{x} +\sin \psi \sin\chi \hat{y}+\cos\chi \hat{z} .
\end{equation}
The orbital velocity of $M_1$ is
\begin{equation}
    \vec{\beta}_H = \beta_H \left(\cos{\left(\frac{2\pi}{P} t\right)}\hat{x} +  \sin{\left(\frac{2\pi}{P} t\right)}\hat{y}\right), 
\end{equation}
in the lab frame. We express the line of sight direction as 
\begin{equation}
    \hat{n}=\sin i \hat{x} +\cos{i} \hat{z},
\end{equation}
where $i$ is the angle between $\hat{n}$ and the jet axis at its base (assumed to be along $\hat{z}$). The middle panel of Fig. \ref{fig:modeldrawing} shows these vectors labeled on the jet. 

The angle between $\hat{r}_0$ and $\hat{n}$ in terms of $\chi$ and $\psi$ is 
\begin{equation}
    \cos\theta_0(\chi, \psi) = \sin i  \sin\chi \cos\psi +\cos i \cos \chi.
\end{equation}
The variation in Doppler beaming at a height $r$ above the black hole is
\begin{equation}
\label{eq:variation_wind}
    \delta D = A_0 \cos{\left(2\pi\left(\frac{t}{P} +\frac{r}{\lambda_h}\right) -\phi_c\right)},
\end{equation}
where 
\begin{equation}
\begin{split}
    &A_0\equiv\frac{\beta_j}{\beta_w}\frac{\beta_H}{\Gamma_j(1-\beta_j\cos\theta(\chi, \psi))^2}\\& \times\sqrt{\sin^2{i}-2\cos\theta_0\sin{i}\sin{\chi}\cos{\psi}+\cos^2\theta_0\sin^2{\chi}},
\end{split}
\end{equation}
and
\begin{equation}
    \tan\phi_c = \frac{\cos\theta_0 \sin{\chi}\sin{\psi}}{\cos\theta_0\sin{\chi}\cos{\psi}-\sin i}.
\end{equation}
The $r$-dependent phase shift in eq. \ref{eq:variation_wind} is caused by the helical structure (shown in the bottom panel of Fig. \ref{fig:modeldrawing}), which propagates with pattern speed $\beta_w$.

\subsubsection{Linear Polarization}
Given that the magnetic field is assumed to be transverse\footnote{Observational justification for our assumption of a toroidal magnetic field is given in \citet{2024Paper2}, as the polarization in PKS 2131 is observed to be along the large-scale jet direction.}, the electric vector should be along $\hat{r}$.  
The electric vector polarization angle (EVPA) and polarization degree (PD) will vary because the angle between $\hat{r}$ and $\hat{n}$ changes. The polarization properties discussed here will be for one specific position ($r, \chi, \psi$) in the jet. Different $(r,\chi,\psi)$ positions will have the same general polarization pattern but be slightly out of phase, leading to some decoherence.

If looking at a ring of magnetic field face on, there is no preferred direction and thus no net polarization over the whole ring; however, when looking at an angle, the observed electric vector will be parallel to the sky-projected jet direction. In the observer frame, the EVPA rotates over the binary period. The observed electric vector is
\begin{equation}
\label{eq:electricvector}
    \vec{e}_{obs}=\frac{\hat{r}-\left(\hat{r}\cdot\hat{n}\right)\hat{n}}{\sqrt{1-\left(\hat{r}\cdot\hat{n}\right)^2}}.
\end{equation}
{The observed PD will be a function of viewing angle. For a transverse field, more edge on vantage points give higher polarization.} Thus, the observed PD is
\begin{equation}
\label{eq:polarizationdegree}
    \Pi_{\rm obs}=\Pi_{\max}\left[1-\cos^2 \theta\right]=\Pi_{\max}\left[1-\left(\hat{r}\cdot\hat{n}\right)^2\right],
\end{equation}
where $\Pi_{\max}$ is {the PD when viewing completely edge on. $\Pi_{\max}$ will be suppressed from the nominal synchrotron polarization limit due to the level of magnetic turbulence present in the emission zone.} 

We decompose eq.~\ref{eq:electricvector} and eq.~\ref{eq:polarizationdegree} into a constant and varying component
\begin{equation}
    \vec{e}_{\rm obs}=\vec{e}_0+\delta\vec{e},
\end{equation}
and 
\begin{equation}
    \Pi_{\rm obs}=\Pi_0+\delta\Pi.
\end{equation}
The non-varying electric vector (which represents the average value) is
\begin{equation}
    \vec{e}_0 = \frac{\hat{r}_0-\left(\hat{r}_0\cdot\hat{n}\right)\hat{n}}{\sqrt{1-\left(\hat{r}_0\cdot\hat{n}\right)^2}},
\end{equation}
while the variation is
\begin{equation}
\label{eq:electrocvectorvar}
    \delta\vec{e} =\frac{\frac{\vec{\beta}_H}{\beta_w}-\left(\frac{\vec{\beta}_H}{\beta_w}\cdot \hat{n}\right)\left(\left(1+\left(\hat{r}_0\cdot\hat{n}\right)^2\right)\hat{n}-\left(\hat{r}_0\cdot\hat{n}\right)\hat{r}_0\right)}{\sqrt{1-\left(\hat{r}_0\cdot\hat{n}\right)^2}}.
\end{equation}
We define the variation in EVPA by
\begin{equation}
\label{eq:EVPAvar}
    \delta\cos \text{PA} =\frac{\vec{e}_0\cdot{\delta\vec{e}}}{\left|\vec{e}_0\right|\left|\delta\vec{e}\right|}.
\end{equation}
To lowest order in $\beta_H/\beta_w$, the variation in PD is 
\begin{equation}
\label{eq:PDvar}
    \delta \Pi = -2\Pi_{\max}\left((\hat{r}_0\cdot\hat{n})(\hat{n}-\left(\hat{r}_0\cdot\hat{n}\right)\hat{r}_0)\right)\cdot\frac{\vec{\beta}_H}{\beta_w}.
\end{equation} 
These vary sinusoidally, and can be out of phase with the flux density variations depending on angular position  ($\chi$, $\psi$) within the jet. We note that eqs. \ref{eq:EVPAvar} and \ref{eq:PDvar} are valid when $\beta_H/\beta_w<i$. When $\beta_H/\beta_w\gtrsim i$, the polarization variation can no longer be treated perturbatively and must be computed by eqs. \ref{eq:electricvector} and \ref{eq:polarizationdegree}.

\subsection{Emission}
 
We integrate over the entire jet to compute the intensity and flux density. The optical depth is $\tau_\nu(r, \chi,\psi)=\int_r^\infty d\tilde r  \cos{\theta(\chi, \psi)}\mu_\nu(\tilde r )$. We assume the source is at redshift $z_S$ and define a source frame frequency $\nu_z=(1+z_S)\nu$ where $\nu$ is the observed frequency. The intensity is 
\begin{equation}
\label{eq:intensity}
    I_{\nu, \Omega}(\chi, \psi, \nu) = (1+z_S)\,d_L^{-2}\int j_{\nu_z, \Omega} e^{-\tau_\nu}\cos\theta(\chi, \psi)dr,
\end{equation}
where $d_L$ is the luminosity distance corresponding to $z_S$ and the integration bounds are $r\in[r_{\min}, r_{\max}]$. {For calculating $d_L$, we assume a flat $\Lambda$CDM universe with $H_0=67.8$ km s$^{-1}$ Mpc$^{-1}$ and $\Omega_m=0.308$.} $r_{\min}$ and $r_{\max}$ are the minimum and maximum radii from the black hole which contribute emission at frequency $\nu$. $r_{\min}$ is set by the radius of the photosphere that is optically thick to frequency $\nu$. $r_{\max}$ is more poorly constrained and may relate to where either the collimating wind ceases or synchrotron emission at frequency $\nu$ can no longer be emitted. We return to $r_{\max}$ in Sec. \ref{sec:PKS2131_model}.

The observed flux density at $\nu$ is
\begin{equation}
\begin{split}
\label{eq:fluxdensity_total}
S_\nu&=(1+z_S)\,d_L^{-2}\iiint j_{\nu_z, \Omega} e^{-\tau_\nu} dV. 
\end{split}
\end{equation}
For volume element $dV=r^2 \sin\chi dr d\chi d\psi$, the integration bounds are $r\in[r_{\min}, r_{\max}]$, $\chi\in[0, \theta_j]$, and $\psi \in[0, 2\pi]$. 

The binary motion causes variation in intensity 
\begin{equation}
\begin{split}
\label{eq:variation_intensity}
 \delta I_\nu &= (1+z_S)\,d_L^{-2}\int\left( \delta j_{\nu_z, \Omega} - j_{\nu_z, \Omega} \delta \tau_\nu\right) e^{-\tau_\nu}\cos\theta dr,  
 \end{split}
\end{equation}
and flux
\begin{equation}
\begin{split}
\label{eq:variation_fluxdensity}
 \delta S_\nu &= (1+z_S)\, d_L^{-2}\iiint\left( \delta j_{\nu_z, \Omega} - j_{\nu_z, \Omega} \delta \tau_\nu\right) e^{-\tau_\nu}dV, 
 \end{split}
\end{equation}
where the integration bounds are the same as eqs. \ref{eq:intensity} and \ref{eq:fluxdensity_total}. Note that since the variation is in $D$, $\delta j_{\nu_z, \Omega}\approx(2+\alpha) \frac{\delta D}{D_0}j_{\nu_z, \Omega}$ and $\delta \tau_\nu\approx(1.5+\alpha) \frac{\delta D}{D_0}\tau_\nu$. 
With these substitutions, we can compute eqs.~\ref{eq:variation_intensity} and \ref{eq:variation_fluxdensity} using eq.~\ref{eq:emissivity} and eq.~\ref{eq:variation_wind} evaluated at observer time $t_o$ (defined such that $t=t_o-r\cos\theta/c$). 

\subsection{Model Properties}
\label{subsubsec:helixmodel}
\begin{figure}
    \centering
    
    \includegraphics[width=\linewidth]{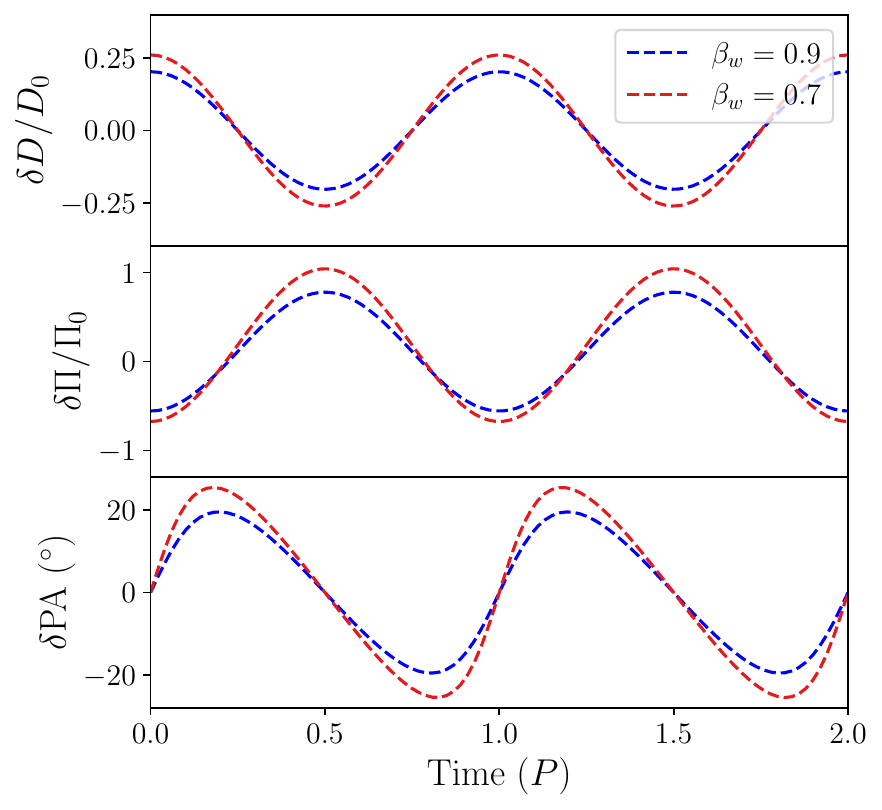}
    \caption{The local fractional variation in Doppler factor $\delta D/D$ (top), PD $\delta \Pi/\Pi_0$ (middle), and EVPA $\delta$PA in degrees (bottom) for choices of $\beta_w$ at a position $(\chi, \psi)=(0,0)$ within the jet. We fix $\beta_H=0.02$, $\Gamma_j=10$, and $i=3.8^\circ$.}
    \label{fig:modelvariation}
\end{figure}
The beaming and polarization variations from a local patch of the jet are given by eqs.~\ref{eq:variation_wind}, \ref{eq:EVPAvar}, and \ref{eq:PDvar}. In Fig.~\ref{fig:modelvariation}, we plot $\delta D/D_0$, $\delta \Pi/\Pi_0$, and $\delta \text{PA}$ for emission from along the jet axis (i.e. $(\chi, \psi)=(0,0)$ so $\hat{r}_0=\hat{z}$) at some $r$. {The precise value of $r$ does not affect the relative phase between $\delta D/D_0$, $\delta \Pi/\Pi_0$, and $\delta \text{PA}$ and only sets the absolute phase.} We set $i=3.8^\circ$ as inferred for PKS 2131. The ratio $\beta_H/\beta_w$ controls the variation because the slower the wind, the more rapidly the jet direction changes. Variation in PD is $\delta\Pi/\Pi_0\approx-2(\beta_H/\beta_w)\cot i\cos(2\pi t/P)$ when $\chi\approx0^\circ$. For this very simple model, note that $\Pi_0=0$ for $i=0^\circ$ (when $\chi=0^\circ$), {corresponding to no preferred field direction}. Consequently, at small $i$ (i.e. when the viewer looks directly down the jet), {$\delta\Pi$ contains all the induced polarization and is not perturbative.} When $\beta_H/\beta_w \lesssim i$ as shown in Fig. \ref{fig:modelvariation}, the PD and EVPA oscillate with orbital period $P$ (like the Doppler factor and flux variation).  When $\beta_H/\beta_w \gtrsim i$, the non-perturbative variation will cause the EVPA to swing by $180^\circ$ with period $\sim P/2$. We note that this is merely the local polarization behavior from an individual spot on the jet. These represent upper limits on the polarization variation observable when integrating over the entire emitting region. In PKS 2131, ALMA polarization data show possible EVPA rotation by $\sim20^\circ$, while the PD varies from $\sim 4\%$ to $8\%$. Max PD is coincident with minimum flux \citep{2024Paper2}. This is compatible with our model if the contribution of several emission zones smears out the variation.

We now discuss the model features when integrating over the whole jet. The absolute flux density depends most strongly on $L_m$ and $L_e$ as well as $r_{\max}$. When integrating over the jet, the amplitude of the sinusoidal variation depends not only on $\beta_H/\beta_w$ but also on $r_{\max}$. Larger $\beta_H/\beta_w$ naturally produces larger amplitude. The relationship between $\beta_w$ and $r_{\max}$ is more subtle, since $\beta_w$ sets $\lambda_h$. Note that when evaluating at observer time $t_o$, the radially dependent phase in eq.~\ref{eq:variation_wind} is
\begin{equation}
        \phi(r)=\frac{2\pi r}{\lambda_h}\left(1-\beta_w\cos\theta\right).
\end{equation}
Integrating eq.~\ref{eq:variation_fluxdensity} to $r_{\max}\gg\lambda_h$ can smear out the variation. Consequently, large $\lambda_h$ leads to stronger flux variation. The final prediction of this model is that sinusoidal variation will be out of phase at different observing frequencies due to the frequency dependent optical depth (i.e. $\tau_\nu\propto\nu^{-\frac{(p+4)}{2}}$). Higher frequencies can thus contribute emission at smaller $r$, causing their light curves to be shifted ahead of lower frequencies. Along with $\beta_w$, $r_{\max}$ determines the precise phase shift trend and is likely related to the extent of the wind collimation or particle acceleration inside the jet. 

The model we sketch here can be readily generalized to more specific observations of these sources (e.g. polarization and multi-wavelength imaging studies) as well as future-identified sources. Targeted numerical simulations will enhance our model.
Relativistic magnetohydrodynamics (RMHD) simulations can provide more realistic descriptions of the jet geometry, magnetic field structure, and wind dynamics, while kinetic simulations can capture the particle acceleration and cooling along the jet.

\section{Interpretation of Observations}
\label{sec:SMBHbinary}
\begin{figure*}
   \centering
   \includegraphics[width=1.0\linewidth]{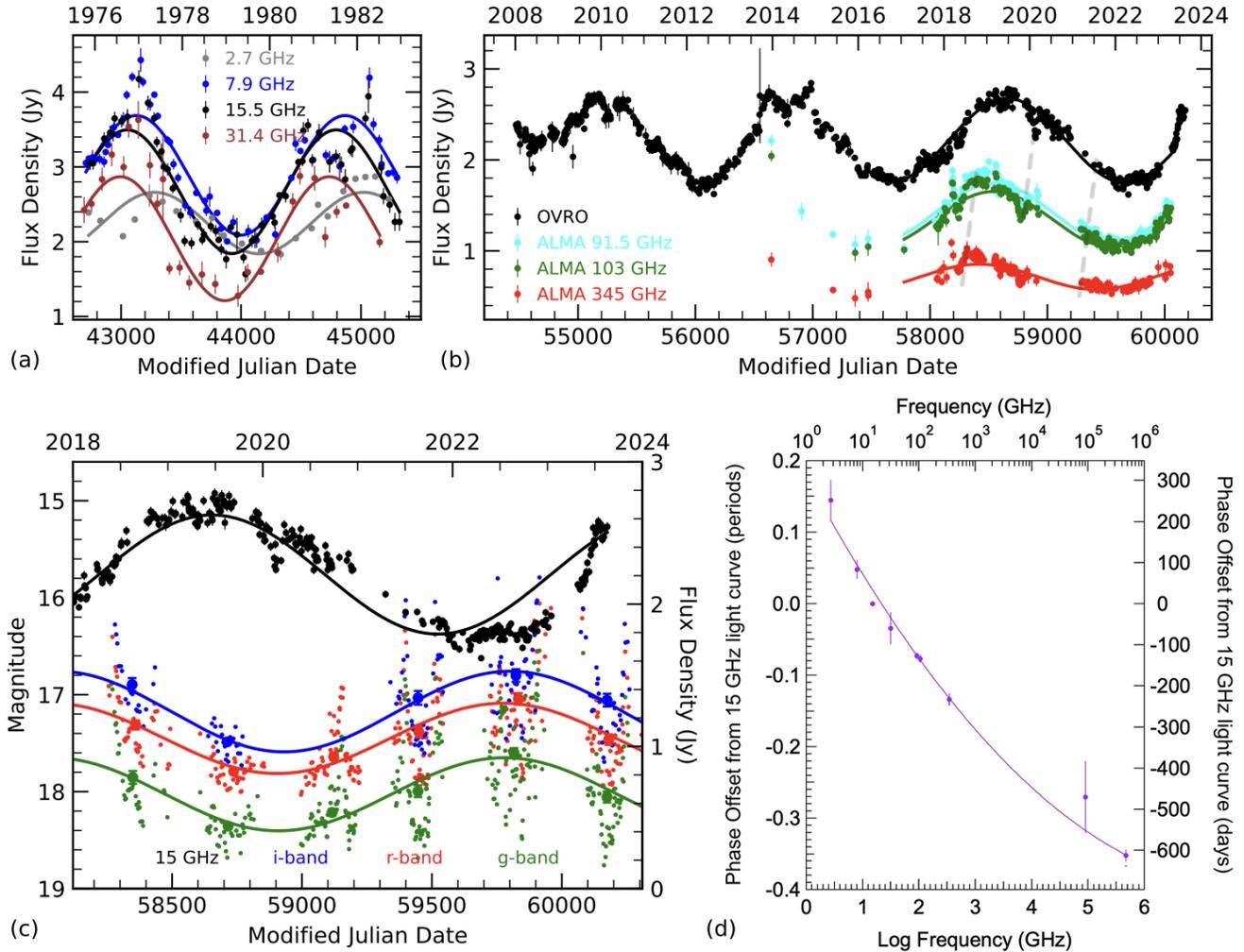}
     \caption{PKS~2131$-$021 light curves at radio, mm, sub-mm,  and optical wavelengths. (a)~Haystack 2.7--31.4 GHz observations \citep{1986AJ.....92.1262O} together with the least-squares sine-wave fits to the data at each of the frequencies. (b)~OVRO 15~GHz light curve plus ALMA light curves at~91.5, 104,  and~345~GHz. (c) Comparison of the $i$, $r$, and $g$-band ZTF optical and OVRO radio light curves of PKS 2131. The curves show the least squares sine wave fits to the corresponding data. (d) The observed phase shifts, relative to the OVRO 15 GHz light curve.  A positive phase shift indicates that the light curve is shifted to a later time than the 15 GHz light curve. The curve shows a quadratic polynomial fit to the phase offset. Reproduced from Fig. 6 in \cite{2024Paper2}.}
         \label{plt:lightcurves3}
\end{figure*}
\subsection{PKS 2131-021}

The $P=1739.8\pm17.4$ day sinusoidal variation in PKS 2131 has $5.06\sigma$ significance thanks to the multi-wavelength observations across multiple epochs  \citep{2024Paper2}.
The most well-sampled light curve (OVRO 15 GHz) has sinusoidal flux density variations with amplitude $\sim0.5$ Jy. There is tantalizing evidence for a higher harmonic with $P/2$ whose amplitude is $\sim12$ mJy.
The fundamental harmonic is significantly identified in 2.7-31.4 GHz Haystack data, 15 GHz OVRO data, 91.5-345 GHz ALMA observations, and ZTF gri data. Fig.~\ref{plt:lightcurves3} shows the light curves in each of the aforementioned bands. Recent ACT light curves complement these datasets in the mm band \citep{2025arXiv250404278H}. The periodicity shuts off from 1984-2003 but returns with the same period and the same phase to within the uncertainties  \citep{2022ApJ...926L..35O}. 
The persistence of the sinusoidal variation is very likely caused by inherent periodicity in the source, consistent with the orbital motion of an SMBHB.

The sinusoidal light curves at low observing frequencies lag behind high observing frequencies. The phase shifts as a function of frequency can be well fitted by a power-law $\delta\phi\propto \nu_o^{-n}$, where $n=0.3$ (See bottom right panel in Fig. \ref{plt:lightcurves3} and discussion in \citealt{2025arXiv250404278H}). 
\subsubsection{Model Application}
\label{sec:PKS2131_model}
\begin{figure}
   \centering
   
   \includegraphics[width=\linewidth]{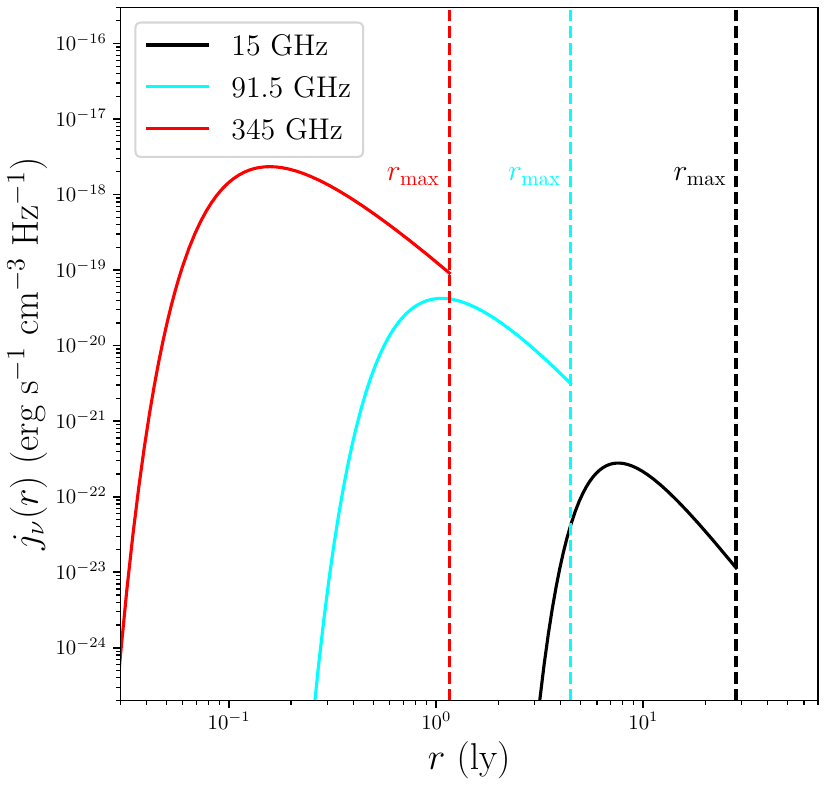}
     \caption{The emissivity as a function of radius $j_{\nu}(r)$ at three frequencies. The emissivity is assumed to be non-zero between $r_{\min}(\nu)$ to $r_{\max}(\nu)$. See text for details.}
         \label{fig:emissivity_radius}
\end{figure}
\begin{figure*}
   \centering
   \includegraphics[width=1.0\linewidth]{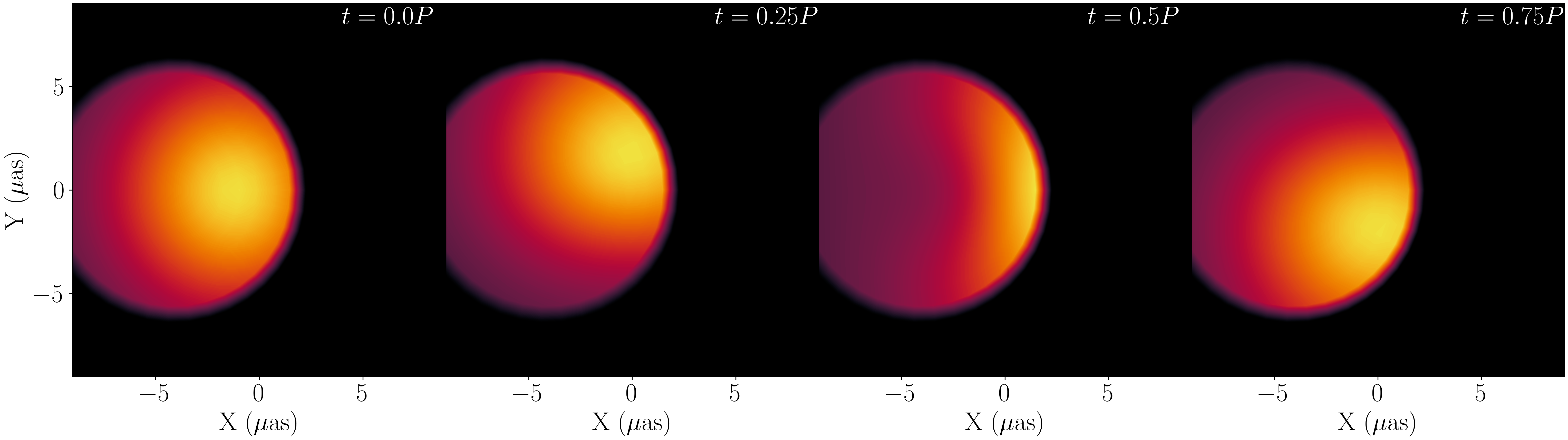}
   \includegraphics[width=0.48\linewidth]{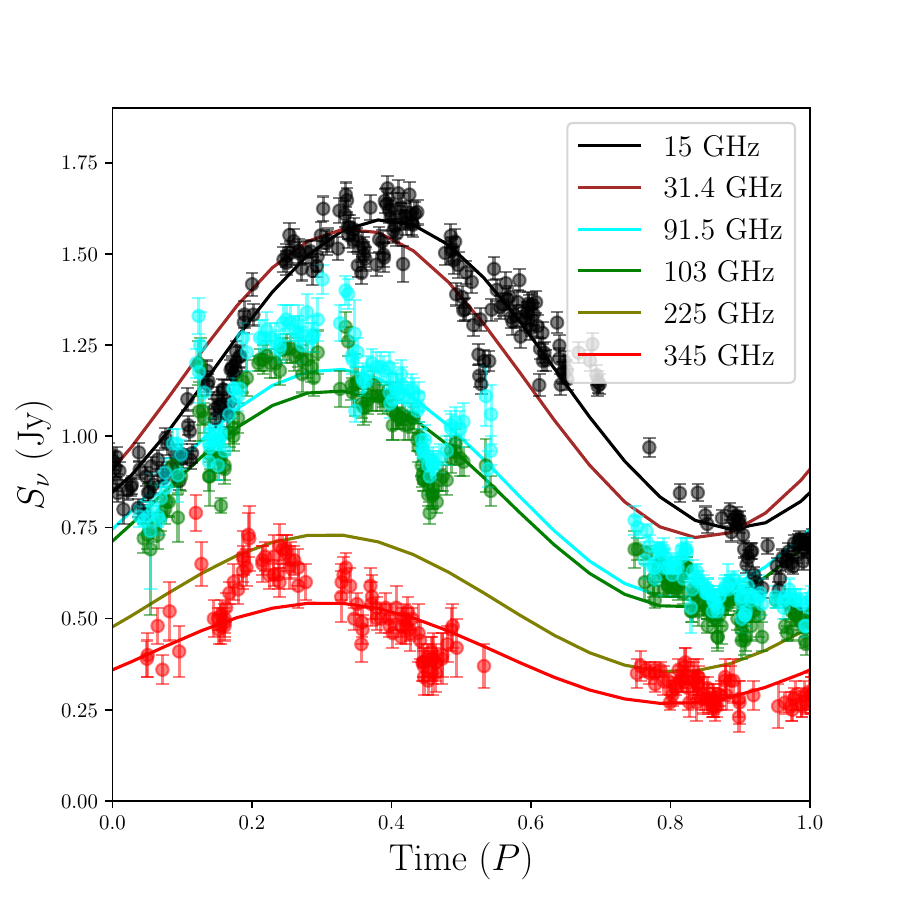}
   \includegraphics[width=0.48\linewidth]{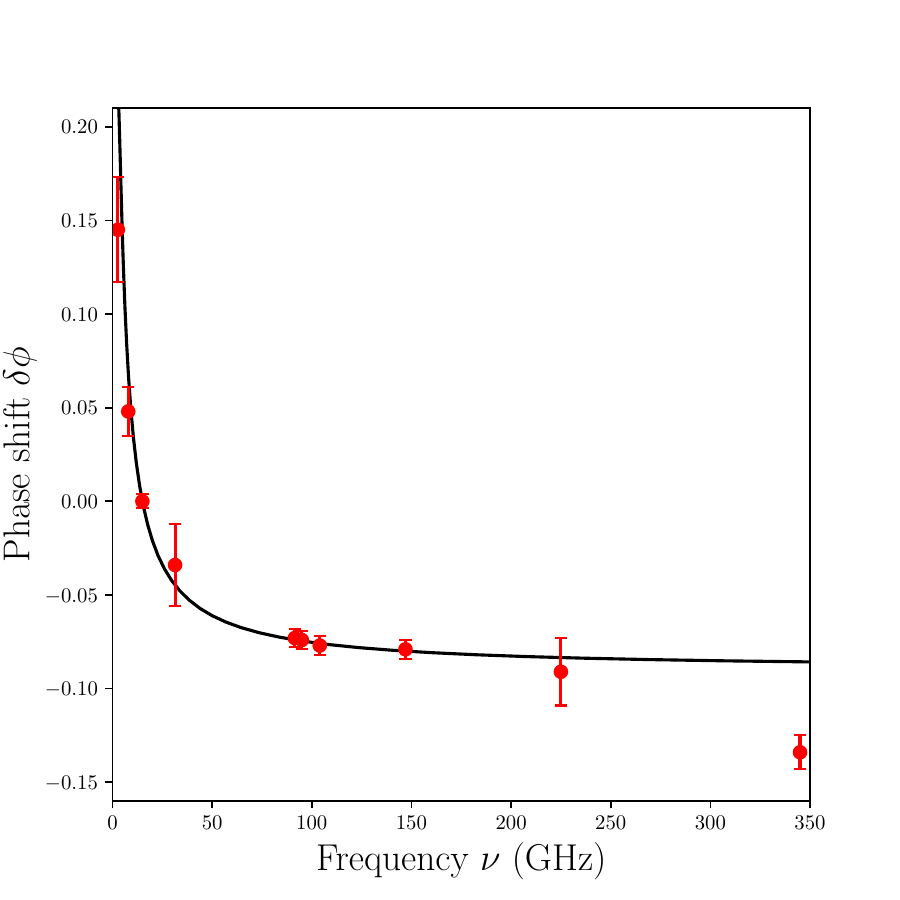}
     \caption{Our model for PKS 2131. (Top) The predicted intensity map at 230 GHz. (Bottom left) The model flux density in a selection of observed bands compared with OVRO and ALMA light curves shifted down by a constant value to match the model. (Bottom Right) Model phase shifts relative to the 15 GHz light compared to the observed phase shifts. Parameters are $\beta_w=0.9$, $\beta_H=0.02$, $\theta_j=0.1$, $\Gamma_j=10$, $r_{\min}$ chosen so that $\tau_\nu(r_{\min})=20$, $r_{\max}=r_{\min}+60 \,(\nu (1+z_S) /15 \text{ GHz})^{-1}$ ly and $i=3.8^\circ$. We assume $L_e=10^{46} \text{ erg s}^{-1}\, (r/r_0)/(1+(r/r_0))$ (where $r_0=5$ ly) and $L_m=10^{46} \text{ erg s}^{-1}\, 1/(1+(r/r_0))$.}
         \label{fig:quivering}
\end{figure*}
We apply our model described in Sec.~\ref{sec:helicaljet} to interpret the observations and reproduce the observed sinusoidal variation in PKS 2131. We choose $i=3.8^\circ$, consistent with VLBI measurements of the large-scale jet \citep{2021ApJ...923...67H}. For the base jet, we take $L_e=10^{46} \text{ erg s}^{-1}\, (r/r_0)/(1+(r/r_0))$ (where $r_0=5$ ly),  $L_m=10^{46} \text{ erg s}^{-1}\, 1/(1+(r/r_0))$, $\theta_j=0.1$, and $\Gamma_j=10$. These high jet powers are required to produce $\sim1$ Jy radio flux at $z_S=1.3$, while the scaling is chosen so that $L_j=L_m+L_e$ remains constant. We assume the particle spectrum, whose normalization evolves along the jet, is initialized with $\gamma_{e,\min}=10$, $\gamma_{e,\max}=300$, and $p=1.2$. We return to the spectra later in the discussion. We choose $\beta_w=0.9$, $\beta_H=0.02$, and $r_{\max}=r_{\min}+60 \,(\nu(1+z_S) /15 \text{ GHz})^{-1}$ ly. Fig.~\ref{fig:emissivity_radius} shows the emissivity as a function of radius for the $15$ GHz, $91.5$ GHz, and $345$ GHz emission. In this model, the average radius of the 15 GHz emission is about 20 ly or 10 helical wavelengths. Without the wind, the phase shifts would require the average radius to be 100-500 ly (depending on the jet bulk Lorentz factor), so the emission regions would be separated by factors of several to 10 times more helical wavelengths.

Fig.~\ref{fig:quivering} shows the results. The top panel shows the theoretical 230 GHz intensity map at different phases in the orbit, while the bottom left and right panels show the computed time varying flux density at several observed radio frequencies, and the phase shifts with frequency, respectively. The theoretical curves shown in the bottom left of Fig.~\ref{fig:quivering} are the total flux density predicted by the model. We subtract a constant flux from each real light curve shown so that the average flux of the model matches the observed average flux. We reproduce the observed variation amplitudes while under-predicting the average flux in most bands. We must subtract $\sim1.1$ Jy from the observed OVRO 15 GHz light curve, $\sim0.55$ Jy from the observed ALMA 91.5 and 103 GHz light curves, and $\sim0.3$ Jy from the observed ALMA 345 GHz light curve. These excess fluxes likely come from non-sinusoidally varying large-scale ($\gtrsim100$ pc) jet components \citep{2022ApJ...926L..35O}. These non-varying components are observed in the VLBI maps and dominate during the epoch when the periodicity shuts off. At these scales, it is likely that the jet no longer remains a coherent helix, so periodicity is smeared out. Modeling the large-scale components in detail is beyond the scope of this paper. 

Note that our chosen $\beta_w$ value and $r_{\max}\propto\nu^{-1}$ scaling allow us to reproduce the observed radio phase shift trend (lower right of Fig. \ref{fig:quivering}). The large value of $\beta_w$ is needed to preserve sinusoidal variation when integrating out to $>10$ ly, suggesting a trans-relativistic wind-jet boundary layer. The predicted trend does not extend to optical, which we do not attempt to explain with this model. The optical emission likely originates much closer to the black hole where physical conditions such as $\Gamma_j$ or $\beta_w$ may differ.

The imposed $r_{\max}\propto\nu^{-1}$ scaling matches both the sinusoidal amplitudes and phase shifts. Concretely, the $r_{\max}\propto\nu^{-1}$ trend corresponds to a frequency dependent $r$ cutoff in $j_{\nu, \Omega}(r)$ and, consequently, a global cutoff in the non-thermal particle spectrum at maximum Lorentz factor $\gamma_{e, \max}\sim 300$ as assumed. {Up to now, we have not made any assumptions about the acceleration mechanism needed to sustain our postulated non-thermal spectrum.}  
Magnetic reconnection efficiently accelerates particles to energies $\gamma\sim\sigma=B^2/4\pi mnc^2$ with a very hard spectrum ($p<2$ as we have assumed) \citep{2014ApJ...783L..21S}. The magnetization $\sigma$ is the ratio of magnetic energy to energy in a particular particle species with mass $m$ and number density $n$. If reconnection is the electron/positron acceleration mechanism in these blazars, this implies $\sigma_e\sim 300$ in the acceleration zone. {In terms of the electromagnetic and radiating particle powers, we have $\sigma\approx L_m/L_e= (r/r_0)^{-1}$, which would suggest the dissipation begins at radii $r\sim r_0/\sigma_e\approx0.02$ ly for the parameters we have assumed. 
At this small radius, the particle population would likely be strongly cooled, so the acceleration zone would need to span a range of radii to sustain the particle population.} {If the magnetic energy is principally dissipated into ions, then acceleration via reconnection can continuously occur at larger radii.} In the emitting regions, the ions, {which we have not otherwise accounted for}, could carry a significant amount of energy, so $\sigma_e$ may remain high even as $\sigma_i$ shrinks ($\sigma_i\approx (m_e/m_i)\sigma_e\sim0.3$). Strong dissipation of the background field and ion mixing can occur near the wind-jet boundary layers, so the observed flux may be dominated by boundary layer emission. If further observational and numerical studies point toward this as well, our model can easily be modified so that $L_e$ depends on $\chi$. {Edge-brightened emission is likely to come from reconnection-accelerated particles, while emission from the center of the jets is more likely to come from shock-accelerated particles. More work will be required to understand particle acceleration in the jets.}

Lastly, the intensity map in Fig.~\ref{fig:quivering}, {which we calculate using eqs. \ref{eq:intensity} and \ref{eq:variation_intensity},} shows spatial shifts in the centroid position over the course of the orbit. At the $\sim2$ Gpc distance of PKS 2131, the centroid shifts by $\sim2$ $\mu$as, which is likely too small to be resolved by the Event Horizon Telescope (EHT); however, a larger mm emission region, extending to $\sim 50$ ly as opposed to $\sim 2$ ly as in our fiducial model, could yield detectable shifts. An EHT campaign showing no variation would thus constrain models of the size of the sinusoidally varying emission region. This example illustrates the characteristics of the direct imaging variation we anticipate. EHT could resolve the shifts in soon to be discovered closer sources obeying this model.

\subsection{PKS J0805-0111}
The $P=1240.7\pm4.6$ day periodicity in the PKS J0805 OVRO light curve has $3.82\sigma$ significance over the observational epoch from 2008-2022 \citep{2025ApJ...987..191D}. The flux density varies by $\sim0.2$ Jy around an average flux density $0.4$ Jy. Contemporaneous ACT light curves at 95 GHz, 147 GHz, and 225 GHz also show sinusoidal variations.
The OVRO light curve lags the higher frequency ACT data just as for PKS 2131 \citep{2025arXiv250404278H}, with $\delta\phi\propto\nu_o^{-0.7}$. The periodicity ceases at MJD 59041 and has not been observed since. Possible periodic gamma-ray flares are observed 238 days out of phase with the three most recent 15 GHz peaks \citep{2025ApJ...987..191D}.

\subsubsection{Model Application}
\begin{figure*}
   \centering
   
   \includegraphics[width=0.48\linewidth]{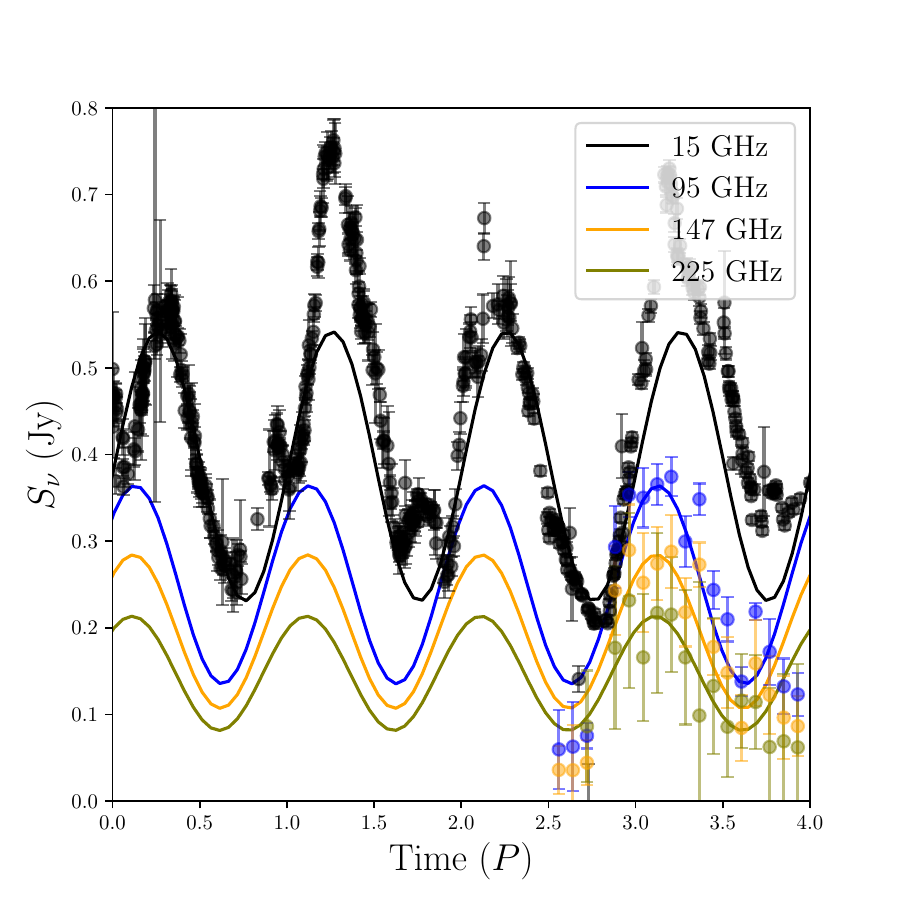}
   \includegraphics[width=0.48\linewidth]{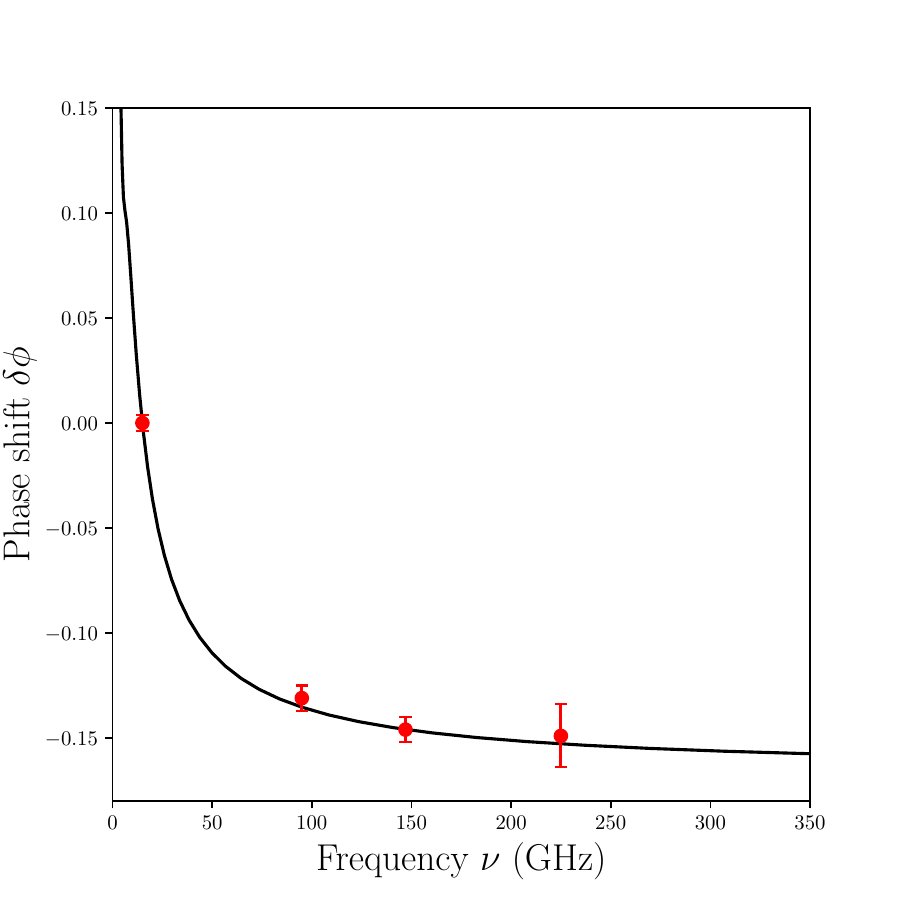}
     \caption{Our model for PKS J0805. (Left) The model flux density in a selection of observed bands compared with OVRO and ACT data. The displayed ACT light curves are each shifted down by a constant flux to match the model. (Right) Model phase shifts relative to the 15 GHz light curve compared to the observed phase shifts. Chosen parameters are the same as Fig. \ref{fig:quivering} except in this model $L_e+L_j=5\times10^{45}$ and $i=5^\circ$.}
         \label{fig:quivering_J0805}
\end{figure*}

We apply our model to PKS J0805 with a similar set of parameters as PKS 2131: $\theta_j=0.1$, $\Gamma_j=10$, $\gamma_{e,\min}=10$, $\gamma_{e,\max}=300$, $p=1.2$, $\beta_w=0.9$, $\beta_H=0.02$, and $r_{\max}=r_{\min}+60 \,(\nu(1+z_S) /15 \text{ GHz})^{-1}$ ly. We use lower jet power $L_e+L_m=5\times10^{45}$ erg s$^{-1}$ and suppose $i=5^\circ$, as there is no VLBI estimated viewing inclination. We show our model results in Fig. \ref{fig:quivering_J0805}. The OVRO peaks have large variability, so we choose to match the lower flux peaks. We do not need to adjust the flux by a constant offset to match the OVRO data, but an additional constant $\sim0.07$ Jy flux is needed to match the ACT data in each band. {The discrepancy in offsets could be due to complexities in emission that we are not modeling. While it is possible to reproduce these features with an ad hoc model, we choose to recover the principal features of the observations with a simple physical description. Note that the ACT light curves correspond to a period where the OVRO flux is high, suggesting that the constant excess needed by the ACT light curves could be the result of a flare in a {non-sinusoidally varying} emission component. Without much VLBI information for this source, however, we cannot otherwise easily distinguish flux from {non-sinusoidal} components.} Nonetheless, our model reproduces the phase shifts in the ACT data well (right panel of Fig. \ref{fig:quivering_J0805}). PKS J0805 light curves at more observing frequencies would be needed to validate the trend and place stronger constraints on these parameters.

\section{Future Observations}
\label{sec:futureobs}
\subsection{Emission Lines}
\begin{figure}
    \centering
    \includegraphics[width=\linewidth]{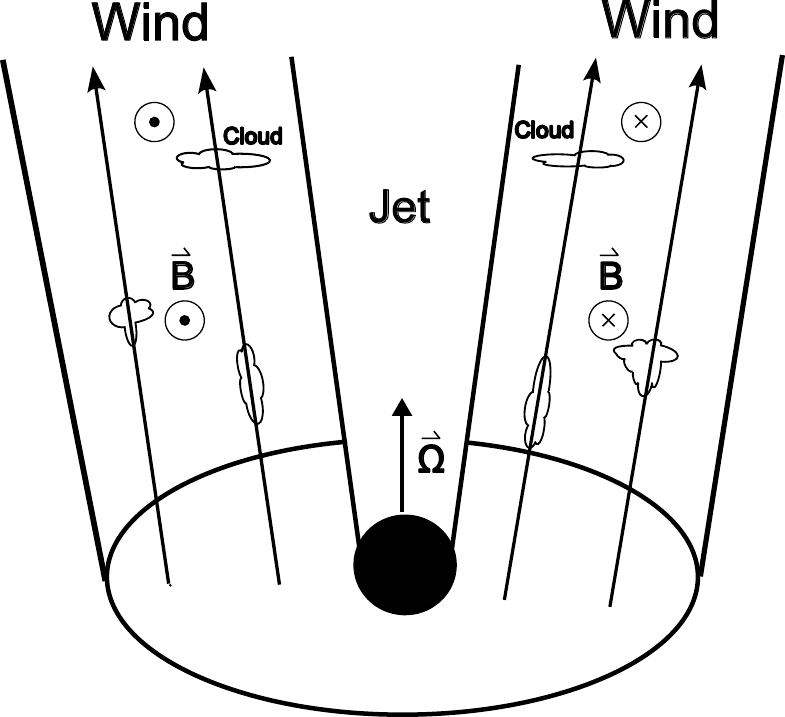}
    \caption{A schematic of the wind-jet complex from the primary spinning black hole. Line emitting clouds travel in the wind along helical magnetic field lines from the disk in the black hole frame and are possibly squished due to a negative velocity gradient. }
    \label{fig:BLR}
\end{figure}
PKS 2131 and PKS J0805 both display emission lines in archival spectra \citep{2006AJ....132....1S, 2008ApJS..175...97H, 2020ApJS..249....3A, 2025arXiv250314745D}.
To date, 4 optical spectra at various epochs have been taken for PKS 2131 and only one has been taken for PKS J0805. More spectra are forthcoming. In both sources, CIII] and MgII lines are visible in the optical band (since $z_S\approx1.3$). These lines likely originate from a broad-line region possibly somewhere in the wind of the jet-producing black hole. There are a vast number of ways these lines could vary as a result of the effects of the binary and wind motion.
For example, in the galactic X-ray binary SS 433, both the luminosity and equivalent width of the prominent H$\alpha$ emission vary by a factor of 3 over the orbit \citep{2004ASPRv..12....1F}. Line widths typically become wider when the accretion disk is oriented edge on. Additionally, H$\beta$ and He I absorption features become weakest when the disk is most face on. Similar orbitally modulated emission and absorption features have been seen in other galactic microquasars \citep[e.g.]{1989MNRAS.238..729F, 2004ASPRv..12....1F}. While provided merely as examples, the aforementioned line behaviors may be present in PKS 2131 and PKS J0805. The phenomenology of line emission in microquasars remains broad, so we can only speculate as to what observations the SMBHB candidates will reveal. 

We propose a model for the broad-line emission variation. In this picture (illustrated in Fig.~\ref{fig:BLR}), the broad line region is composed of cold, dense clouds injected into the wind by the disk \citep{1992ApJ...385..460E}. We envision clouds following helical magnetic field lines in the wind {with speeds which are a function of cylindrical radius from the jet axis}. In the black hole frame, the clouds have velocity 
\begin{equation}
    \vec{v}_{cl}(\rho, \Phi)=v_{cl}(\rho)\left(\sin\Psi\cos\Phi\hat{x}+\sin\Psi\sin\Phi\hat{y}+\cos\Psi\hat{z}\right)\\
\end{equation}
where $\rho$ is the cylindrical radius, $\Phi$ is azimuthal position, and $\Psi$ is the opening angle of the field lines. Assuming the azimuthal motion is tied to the disk rotation, the cloud speed is
\begin{equation}
    v_{cl}(\rho)=v_{cl,0}\sqrt\frac{\rho_0}{\rho}.
\end{equation}
For simplicity, we imagine the clouds are distributed {uniformly in height above the black hole along the $\hat{z}$ axis,  non-uniformly in cylindrical radius (in the $x$-$y$ plane), and uniformly in $\Phi$, although it is likely that the real spatial distribution of clouds is more complicated.} For this illustrative model, we suppose the clouds are distributed in cylindrical radius as
\begin{equation}
    f(\rho)=2\frac{\rho}{\rho_0^2}\exp\left(-\frac{\rho^2}{\rho_0^2}\right).
\end{equation}
Thus, $v_{cl,0}$ roughly corresponds to the average cloud speed. These choices are merely demonstrative and can easily be modified. 

\begin{figure}
    \centering
    \includegraphics[width=\linewidth]{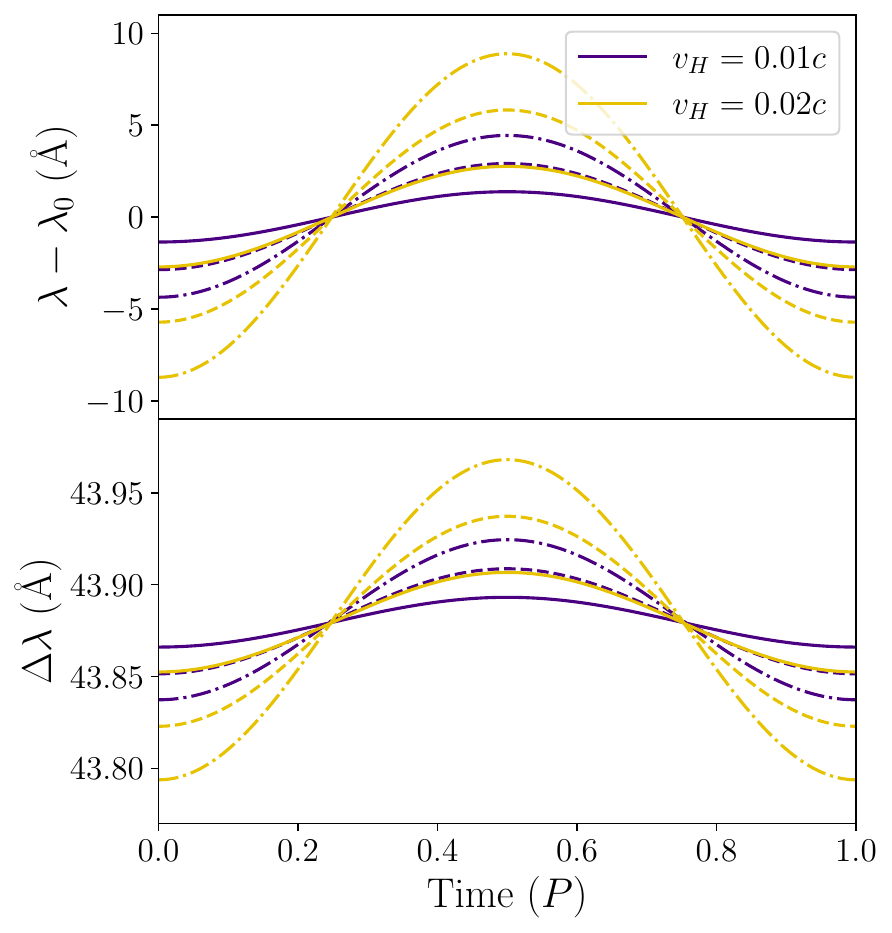}
    \caption{Variation in line emission from our illustrative model. We set $v_{cl,0}=1000$ km s$^{-1}$ and $\Psi=30^\circ$. We plot $i=1.8^\circ$ with solid lines, $i=3.8^\circ$ with dashed lines, and $i=5.8^\circ$ with dot-dashed lines for two $v_H$ values.}
    \label{fig:BLR_variation}
\end{figure}

In the observer frame, the cloud velocity is
\begin{equation}
  \vec{v}_{cl}^\prime=\vec{v}_{cl}+\vec{v}_{H}.
\end{equation}
Lines emitted by the distribution of clouds will have central wavelength {determined by the average Doppler shift from the distribution of clouds}
\begin{equation}
    \frac{\lambda(t)}{\lambda_0}=\frac{1}{2\pi}\int \left(1-\frac{\vec{v}_{cl}^\prime\left(\rho, \Phi, t\right)}{c}\cdot\vec{n}\right)f(\rho)\text{d}\rho\text{d}\Phi.
\end{equation}
{The line width will be the standard deviation of the Doppler shifts}
\begin{align}
    \frac{\Delta \lambda(t)}{\lambda_0}&=\nonumber\\&\sqrt{ \frac{1}{2\pi}\int\left(1-\frac{\vec{v}_{cl}^\prime\left(\rho, \Phi, t\right)}{c}\cdot\vec{n}\right)^2f(\rho)\text{d}\rho\text{d}\Phi-\frac{\lambda(t)^2}{\lambda_0^2}}.
\end{align}
We show the variation in central wavelength and width for a line with nominal wavelength $\lambda_0=4400$ $\mathrm{\AA}$, corresponding to the semi-forbidden $1909$ $\mathrm{\AA}$ CIII] line at $z_S=1.3$ in Fig.~\ref{fig:BLR_variation}. The line centers vary by $\sim2-5$ $\mathrm{\AA}$ at the nominal $i=3.8^\circ$ for PKS 2131 depending on the orbital velocity, while the variation increases for higher inclinations. These small shifts can be measured with several observations per orbit over the course of multiple orbits. Together with the amplitude of the flux variation, these shifts reveal the orbital speed and consequently the component masses. This simple example is meant to be illustrative. The real variability will be more complicated. Only more observations will tell.

A lack of any unique line variation, however, would severely hamper the binary hypothesis. Although possible, it seems unlikely that the jet is produced by a lighter secondary while the broad line region is confined to a slowly moving primary. More likely, a lack of line variation would signal that the periodic behavior is entirely a feature of the jet. This leads to the equally interesting possibility that some global jet mode can disappear and return with $\sim10$ cycles of coherence.

 \subsection{Gravitational Waves}
 \label{subsec:GWs}
 Given their few year periods, PKS 2131 and PKS J0805 likely contribute to the PTA-detected GW background. The GW amplitude and inspiral time depend on the black hole masses, which are at present unknown.  $M_1\sim10^6-10^9$ M$_\odot$ give GW characteristic strains $\sim10^{-17}-10^{-14}$. SMBHBs with $M_1\approx10^8$ M$_\odot$ represent the lightest systems detectable by the current international network of PTAs at these redshifts, and correspond to our fiducial $\beta_H\approx0.02$. These sources are expected to merge in $10^4-10^5$ yr, while $10^7$ $M_\odot$ sources (corresponding to $\beta_H\approx0.01$) will likely merge in $10^6-10^7$ yr. PKS 2131 and PKS J0805 should both merge well within the age of the universe, but are far enough from merger to remain continuous non-evolving GW sources. These and future radio-identified binaries provide definitive locations on the sky and GW periods, as well as binary phase, orbital velocity, and observing inclination constraints --- all important priors for targeted GW searches in PTA data. This motivates searches through the radio PTA data as well as available Fermi gamma-ray pulsar timing data, which can also constrain nHz GW signals \citep{2022Sci...376..521F}. NANOGrav PTA data has begun to place upper limits on these systems' chirp masses ($\lesssim10^{10}$ M$_\odot$) \citep{2025arXiv250816534A} based on their sky locations and periods. These searches can be enhanced by adding more priors from our modeling as well as longer observing baselines. Convincing evidence of GWs from even one source would represent a major discovery.

\section{Conclusions}
\label{sec:conclusion}
SMBHBs should be common. Galaxies are observed to merge, and it is widely accepted that large galaxies, in particular, form through sequences of mergers. Given that these galaxies host supermassive black holes (SMBHs), SMBHBs should naturally form in their centers through dynamical friction. SMBHBs are the most natural source of nHz GWs. The remarkable lack of direct evidence for these systems represents a new ``missing mass" problem and motivates continued observational searches coupled with followup theoretical modeling to understand their signatures and their environments. 

We have developed a semi-analytic model for the emission produced by jetted SMBHBs. PKS 2131 and PKS J0805 provide an opportunity to test models of relativistic jets in an unprecedented way. The flux density variations and phase shifts directly probe the internal structure of the jets. The phase shifts motivate the presence of a sub-relativistic wind which controls the direction of the radio emitting jet. 
This picture allows us to infer jet physics including high $10^{45}-10^{46}$ erg s$^{-1}$ jet powers, an effective speed of collimating wind-jet boundary $\beta_w\approx0.9$, as well as hints of the acceleration mechanisms for the radiating particles. Our models reproduce the observed phenomena well with rather similar parameters for the two sources, suggesting consistent features within this new class of systems. 
Simulations of a helical jet collimated by a magnetized outflow can test the stability of this model and produce a more physically motivated emissivity. 

{This model, while sufficient for our current need, neglects jet structure and is limited to small perturbations. 
An improved model will need to include three additional features:
\begin{enumerate}
\item The jet emissivity and absorption coefficient vary as a function of angular position ($\chi$,$\psi$). In particular the relativistic electron density can be larger near the jet surface --- the sheath --- and at its leading edge where it will interact most strongly with the hypothesized confining wind.
\item The particle acceleration and synchrotron cooling vary as a function of radius along the jet.
\item The effective wind-jet boundary velocity $\beta_w$ may vary with radius. However, its temporal variation is strictly periodic. 
\end{enumerate}
Models with these features can help separate the kinematical from the dynamical contributions to the flux variation and will be discussed in future work. This may also be useful for interpreting RMHD simulations.}

Of the two sources, PKS 2131 remains the stronger SMBHB candidate. 
Recent ACT observations have strengthened the case for PKS J0805. The few yr periods make these sources some of the most promising individually identifiable nHz GW emitting candidates.
Ultimately, more data is required to determine if either is indeed a binary as well as to probe the putative wind. Near-IR observations could bridge the gap between radio and optical to extend the phase shift trend. High quality spectra will be key, as they would contain imprints of the orbital velocities and wind velocities and show variability \citep[e.g.][]{2005ApJ...622L.129B}. A lack of spectral variation would suggest the SMBHB hypothesis is severely deprecated.

Long-term blazar monitoring with OVRO has revealed the existence of this class of possible SMBHBs. Simons Observatory and other mm and cm instruments continuously monitoring the sky may find more sources with similar phenomenology. The periods seen in PKS 2131 and PKS J0805 have been both predictive and retrodictive, {as the sinusoidal variation has been corroborated by both new and preexisting data.} As PKS 2131 has proven, extant archival data has the ability to make less significant periodicities more significant. Additionally, while the shifts in the spatial position centroid of these sources may be too small to be resolved by EHT, variation in more nearby sources, which may soon be found, could be directly imaged. Continued monitoring and additional multi-wavelength observing programs can secure these sources as binaries and identify more SMBHB candidates with relativistic jets.

\section*{Acknowledgments}
The authors thank Alisa Galishnikova,  Martijn Oei, Timothy J. Pearson, Viraj Manwadkar, Jack Dinsmore, Roger W. Romani, Emmanuele Sobacchi, Alexander Philippov, Lorenzo Sironi, and Paulo Coppi for useful discussions. The authors also thank the anonymous referee for useful feedback which improved this manuscript. This work was supported in part by a grant from the Simons Foundation (00001470, R.B., A.G.S., N.G.).
A.G.S. acknowledges the support of the Stanford University Physics Department Fellowship, the National Science Foundation Graduate Research Fellowship, and a Giddings Fellowship at the Kavli Institute for Particle Astrophysics and Cosmology at Stanford. The work at the Owens Valley Radio Observatory is supported by NSF grants AST2407603 and AST2407604. We thank the California Institute of Technology and the Max Planck Institute for Radio Astronomy for supporting the  OVRO 40\,m program under extremely difficult circumstances over the last 8~years in the absence of agency funding. Without this private support these observations could not have been made.  Prior to~2016, the OVRO program was supported by NASA grants \hbox{NNG06GG1G}, \hbox{NNX08AW31G}, \hbox{NNX11A043G}, and \hbox{NNX13AQ89G} from~2006 to~2016 and NSF grants AST-0808050 and AST-1109911 from~2008 to~2014.

\bibliography{refs}{}
\bibliographystyle{aasjournal}



\end{document}